\pdfoutput=1
\documentclass[journal]{IEEEtran}
\usepackage{graphicx}
\usepackage{adjustbox}
\usepackage{cite}
\usepackage{array}
\usepackage{dblfloatfix}
\usepackage{float}
\usepackage{amssymb}
\usepackage{multirow}
\usepackage{tikz}
\usepackage{tabularx}
\usepackage{booktabs}
\usepackage{adjustbox}
\usepackage{balance}
\usepackage{placeins}
\usepackage{afterpage} 
\usepackage{makecell}
\usepackage[symbol]{footmisc} 
\usepackage{enumitem}

\usepackage{algorithm}
\usepackage{algorithmic}

\usepackage{xcolor}
\usepackage{soul}
\sethlcolor{yellow}

\usepackage{cuted} 
\usepackage{url}
\usepackage{float}

\ifCLASSINFOpdf
\else
\fi

\usepackage{amsmath} 

\IEEEoverridecommandlockouts
\IEEEpubid{\begin{minipage}[t]{\textwidth}\ \\[9pt]
       \centering\normalsize{This work has been submitted to the IEEE for possible publication. \\ Copyright may be transferred without notice, after which this version may no longer be accessible.}
\end{minipage}}

\begin{document}

\title{ADiP: Adaptive-Precision Systolic Array for Matrix Multiplication Acceleration}

\author{Ahmed~J.~Abdelmaksoud, 
        Cristian Sestito, 
        Shiwei Wang, 
        Themis~Prodromakis
\thanks{This work was supported by EPSRC FORTE Programme (Grant No.
EP/R024642/2), the RAEng Chair in Emerging Technologies (Grant
No. CiET1819/2/93), and EPSRC AI Hub for Productive Research and Innovation in eLectronics (APRIL) (Grant No. EP/Y029763/1)}

\thanks{Ahmed J.Abdelmaksoud, Cristian Sestito, Shiwei Wang and Themis Prodromakis are with the Centre for Electronics
Frontiers, Institute for Integrated Micro and Nano Systems, School of Engineering, The University of Edinburgh, EH9 3BF, Edinburgh, United Kingdom
(e-mails: a.j.abdelmaksoud@ed.ac.uk; csestito@ed.ac.uk; shiwei.wang@ed.ac.uk; t.prodromakis@ed.ac.uk).}}%

\maketitle

\begin{abstract}
Transformers are at the core of modern AI nowadays. They rely heavily on matrix multiplication and require efficient acceleration due to their substantial memory and computational requirements. Quantization plays a vital role in reducing memory usage, and can be exploited for computations by designing reconfigurable architectures that enhance matrix multiplication by dynamically adjusting the precision.
This paper proposes ADiP, a novel adaptive-precision systolic array architecture designed for efficient matrix multiplication acceleration. 
The proposed architecture consists of $N$ × $N$ reconfigurable processing elements (PEs), along with shared shifters and accumulators. 
ADiP supports multiple computation modes, including symmetric single-matrix multiplication as well as asymmetric multi-matrix multiplication with a shared input matrix, thereby improving data reuse and PE utilization. 
By adapting to different precisions, ADiP achieves up to 4× higher throughput and up to 4× higher memory efficiency. 
Analytical models are developed for ADiP architecture, including latency and throughput for different architecture configurations. A comprehensive hardware design space exploration is demonstrated using commercial 22nm technology. 
Furthermore, ADiP is evaluated on different Transformer-based workloads from GPT-2 medium, BERT large, and BitNet-1.58B models, delivering total latency improvement up to 53.6\%, and total energy improvement up to 24.4\% for attention workloads in BitNet-1.58B model. 
At a 64×64 size with reconfigurable 4,096 PEs, ADiP achieves a peak throughput of 8.192 TOPS, 16.384 TOPS, and 32.768 TOPS for 8bit×8bit, 8bit×4bit, and 8bit×2bit operations, respectively.

\end{abstract}

\begin{IEEEkeywords}
Hardware Acceleration, Matrix Multiplication, Reconfigurable Computing, Systolic Arrays, Spatial Architectures.
\end{IEEEkeywords}

\IEEEpeerreviewmaketitle

\section{Introduction}

\IEEEPARstart{T}{ransformers} have emerged as the backbone of numerous state-of-the-art systems in natural language processing, computer vision, and multimodal learning, owing to their scalability and capacity for capturing long-range dependencies \cite{vaswani2017attention}. Their effective and scalable architectures have enabled the development of large language models (LLMs) that demonstrate noticeable capabilities in text generation, reasoning, and few-shot learning \cite{minaee2024largelanguagemodelssurvey, raiaan2024review}. However, this progress comes with substantial challenges. The rapid scaling of model size from millions to hundreds of billions parameters has increased the memory requirements, leading to considerable computational costs during both training and inference \cite{naveed2025comprehensive, kim2023full}. The deployment of Transformers, particularly for edge and real-time applications, is constrained by high per-token latency and energy overheads, owing to dense matrix multiplications and memory bottlenecks. These constraints have prompted a surge in research on efficient model execution through compression, quantization, and hardware-aware design.

Quantization has become a pivotal strategy for reducing the inference cost of Transformers and LLMs (Transformer-based models), enabling deployment in memory-constrained and latency-sensitive applications. Two primary approaches allow models to operate at reduced bit-widths while preserving accuracy: quantization-aware training (QAT) and post-training quantization (PTQ) \cite{jacob2018quantization, tang2024survey}. PTQ offers simplicity and speed by calibrating pre-trained weights, while QAT integrates quantization into the training loop, often yielding superior robustness under aggressive compression. 
Hugging Face supports these workflows through frameworks such as Optimum for 8-bit QAT/PTQ pipelines \cite{huggingface2023quant}, while BitsAndBytes extends support to 4-bit and 8-bit inference via QLoRA \cite{dettmers2023qlora}. For lower-bit inference, AutoGPTQ provides highly optimized GPTQ quantization with 2-bit and 4-bit support for hardware-aware deployment \cite{frantar2022gptq}. 
Moreover, recent models such as BitNet demonstrate the feasibility of 1-bit and ternary models, achieving high performance with drastically reduced compute demands compared to conventional full-precision models \cite{diao2023bitnet}. 
Additionally, DeepSeek LLM combines architectural optimizations with reduced precision achieving significantly lower memory and inference costs \cite{liu2024deepseek}.
Together, these tools and models showcase the growing maturity of quantization as a foundational technique for sustainable Transformers and LLMs deployment.

Conventional Von Neumann architectures increasingly fall short in meeting the computational demands of large-scale AI workloads, primarily due to the inherent memory and data movement bottlenecks. Spatial architectures, particularly systolic arrays (SAs), have emerged as promising hardware accelerators to address this limitation \cite{kung1982systolic}. SAs employ a grid of PEs to enhance data locality, with each performing core arithmetic operations such as multiplication and accumulation, coupled with register storage. These arrays improve computational efficiency by maximizing the number of operations per memory fetch, significantly reducing off-chip memory accesses. Data propagates in a rhythmic, wave-like manner across the array, synchronizing with boundary PEs via first-in-first-out (FIFO) buffers. This design naturally promotes data reuse among PEs, particularly beneficial for matrix multiplications \cite{hu2018systolic}. As a result, SAs are increasingly being adopted in modern AI accelerators for their ability to align closely with the dataflow patterns of deep learning workloads.

In the spectrum of hardware acceleration of Transformers, the interplay between quantization techniques and adaptive-precision and reconfigurable architectures maximize computational throughput as well as energy efficiency. These approaches offer the flexibility to dynamically adjust the numerical precision at runtime based on the computational sensitivity of each model component, particularly relevant in multi-head attention (MHA) layers, where the significance of activations varies across stages, heads, and layers \cite{rakka2025mixed, behdin2023quantease}. By tailoring the bit-width per head or layer, systems can minimize the precision without reducing the model performance. 
Consequently, adaptive-precision architectures can deliver higher gains in energy and area efficiency while maintaining high throughput for quantized Transformer-based workloads \cite{ibrahim2022taxonomy}.
Current adaptive-precision architectures cannot maximize throughput and memory efficiency gains in reduced precision scenarios, and a gap persists to provide an efficient hardware architecture for accelerating quantized Transformer-based workloads. 

In this paper, we present ADiP, a novel adaptive-precision systolic array architecture, designed for accelerating matrix multiplication in Transformer-based models. The main contributions of this work are highlighted as follows:
\begin{itemize}
  \item The proposed ADiP architecture incorporates $N$ × $N$ reconfigurable PEs along with shared shifters and accumulators acroess each PE column, and eliminates input and output synchronization FIFOs required by conventional weight-stationary (WS) architecture. 
  \item The proposed ADiP architecture support multiple computation modes, including symmetric single-matrix multiplication and asymmetric multi-matrix multiplication with a shared input matrix. ADiP also adapts to different precisions, such as 8bit×8bit (8b×8b), 8bit×4bit (8b×4b), and 8bit×2bit (8b×2b). 
  \item Enhanced PE utilization and data reuse, achieved by eliminating synchronization FIFOs, keeping weights stationary, and sharing input matrix to multiple weight matrices in asymmetric matrix multiplication, yielding up to 4× higher memory efficiency.
  \item Analytical modeling is developed for latency and throughput for different architectural configurations.
  \item A comprehensive hardware design space exploration is demonstrated using commercial 22nm technology, achieving up to 4× higher throughput for quantized workloads, with different architecture configurations from 4×4 to 64×64. 
  \item ADiP is evaluated on different attention workloads from GPT-2 medium, BERT large, and BitNet-1.58B Transformer-based models, delivering latency improvement up to 53.6\%, and energy improvement up to 24.4\% using attention workloads from BitNet-1.58B model. 
\end{itemize}

This paper is organized as follows: Section II discusses the related work and background. Section III presents reconfigurable PE architecture. Section IV presents ADiP architecture. Section V shows hardware design space exploration, evaluation, and results. Finally, section VI concludes the work.

\section{Related Work \& Background}

This section presents a literature review about adaptive-precision architectures followed by a background on matrix multiplication stages in attention layers in addition to block matrix multiplication.

\subsection{Related Work}

Reconfigurable and adaptive-precision architectures have emerged as key enabler to balance performance, energy efficiency, and flexibility in accelerating deep learning workloads. Among these, systolic arrays stand out due to their inherent data reuse capabilities and spatial parallelism, which are highly effective for matrix multiplication, the core operation in Transformer models. 
Recent research has investigated dynamic precision scaling within systolic array architectures, allowing the run-time adjustment of operand bitwidths to match the varying precisions across different workloads. 

Prior research has primarily focused on developing adaptive-precision PE. 
The architectural design includes bit-serial architectures \cite{lascorz2024atalanta, judd2016stripes, umuroglu2018bismo} and multiplication decomposition architectures \cite{peltekis2023arrayflex, gope2020high, li2022low, li2023precision, liu2025multi, yangbitsystolic, dnpu2017, lee2018unpu, yang2022dtqatten}.
Data-gating is the most widely used method because it simply gates the MSBs to zero \cite{camus2019review}. However, the gated logic cannot be exploited for higher throughput, and its area becomes unutilized. 
Another approach is dynamic voltage-accuracy-frequency scaling (DVAFS), which enables various precision computations and increases parallelism by reusing full-adder cells that are inactive at scaled precision \cite{camus2019review}. However, it does not account for variable inputs. 
Divide-and-conquer (D\&C) PE is another approach, wherein full-precision multiplier is built from shifted binary additions of partial products \cite{dnpu2017}. The intermediate sums correspond directly to the results of the scaled multiplication. The first operand is common to all subword computations, allowing parallelization when one operand has to remain at full-precision. The second operand is split into parallel subwords, doubling the number of operations per cycle for lower-precision computation.
Bit-serial architectures perform matrix multiplication through bit-wise computations.
In this approach, bit-serial multiply-accumulate (MAC) receives weights through 1-bit iterations, while activations are sent bit-wise or in a parallel manner \cite{umuroglu2018bismo}.
These architectures are useful for quantized workloads with variable precision; however, their performance is degraded by additional energy consumption overhead and design complexity.

An orthogonal research effort has concentrated on improving the micro-architecture level. 
DTATrans is designed based on a variable-speed systolic array (VSSA) and output-stationary (OS) dataflow \cite{yang2023dtatrans}. It dynamically quantizes different tokens with mixed bit-levels of 0, 4, and 8 bits, achieving 1.3 TOPS with an energy efficiency of 1.62 TOPS/W and an area efficiency of 0.98 TOPS/mm$^2$. However, the architecture performs only symmetric multiplication, and it is based on OS dataflow, which requires input/output synchronization FIFOs.  
PD-FSA is a flexible systolic array architecture with DVAFS PEs, synthesized using Xilinx Vivado at a frequency of 100 MHz \cite{so2022efficient}. It allows 2, 4, and 8 bits symmetric operations. 
BISMO is a bit-serial matrix multiplication architecture that scales its precision \cite{umuroglu2018bismo}. BISMO achieves a peak performance of binary 6.5 TOPS with an energy efficiency of up to binary 1.4 TOPS/W on a PYNQ-Z1 FPGA. Although the same hardware can be utilized for a range of different precisions, it requires higher computational latency and energy overhead.
In addition, multi-fused multiply-accumulate (MFMA) architecture is a scalable 3D systolic architecture with floating-point MAC units \cite{liu2025multi}. The MFMA scheme instantiates multiple MAC units with fused MAC operations. 

\subsection{Matrix Multiplications In Transformers}

Transformer workloads are becoming increasingly massive and heavily dependent on matrix multiplication, especially in MHA and feed-forward networks (FFN) layers. MHA is the core component of Transformer models that captures complex dependencies in data \cite{vaswani2017attention}. MHA captures diverse representational subspaces by involving multiple heads, allowing the model to understand relationships across different perspectives simultaneously.

\begin{figure}[t]
  \centering
  \includegraphics[width=\linewidth, keepaspectratio]{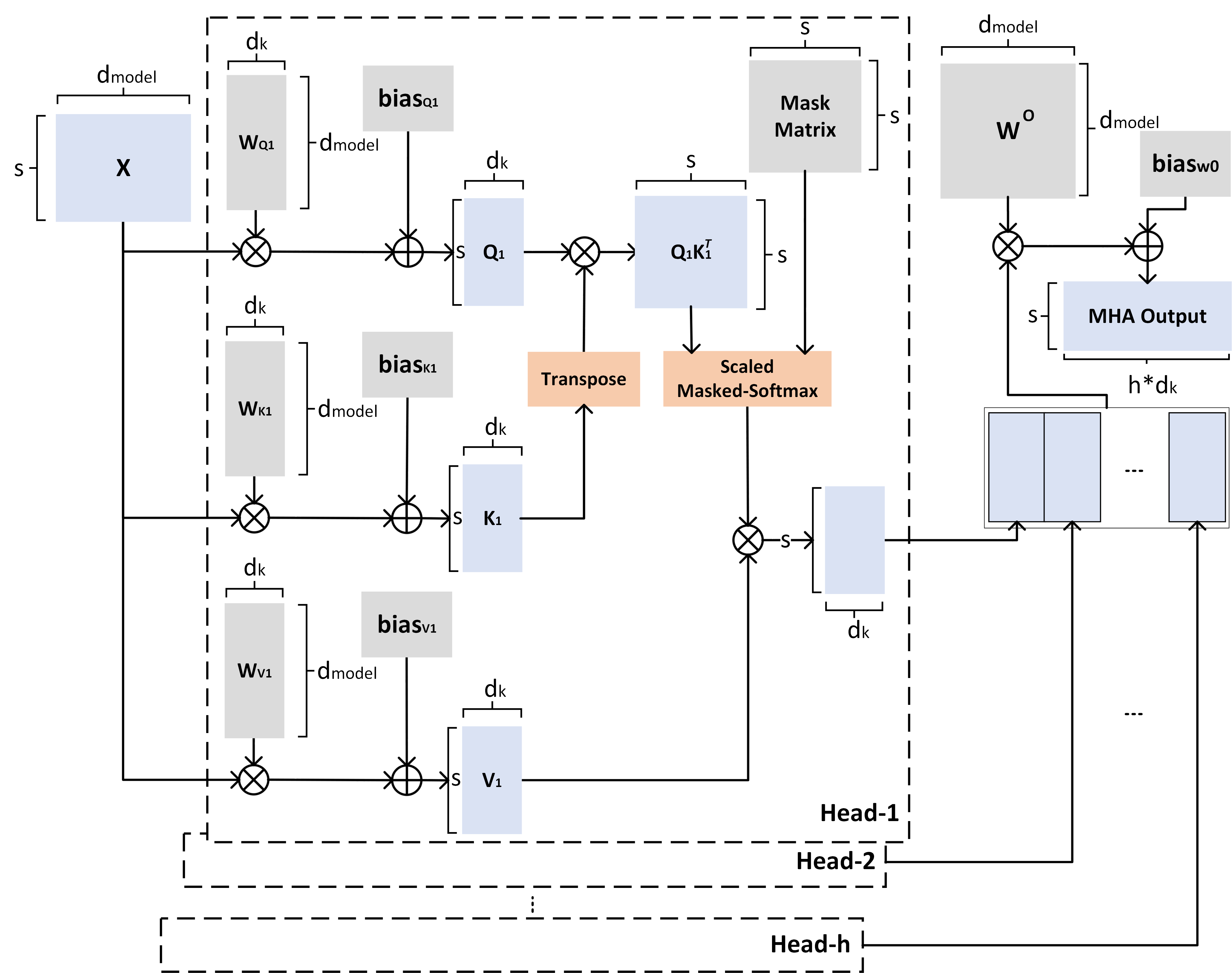} 
  \caption{Matrix multiplication stages for multi-head attention layers.}
  \label{Fig.1}
\end{figure}

Fig. \ref{Fig.1} shows matrix multiplication workloads and their dimensions for MHA, in terms of input sequence length (\(s\)), model size (\(d_{\text{model}}\)), and head size (\(d_k\)). Matrix multiplication in MHA involves four main stages: QKV projections of inputs, attention score, attention output, and output projection.
First, the input (\( X \)) is first projected in queries (\( Q_i \)), keys (\( K_i \)), and values (\( V_i \)) per each head \( i \) using weight matrices. In the second stage, the attention scores (\( S_i \)) for each head are computed by taking the scaled dot product of queries (\( Q_i \)) and transposed Keys (\( K_i \)), followed by a softmax normalization.  In the third stage, the attention scores (\( S_i \)) are multiplied with values (\( V_i \)), producing the attention output (\( Attn_i \)) for each head. The outputs of all heads are concatenated into a single matrix and applied to the output projection matrix (\( W^O \)). 

Matrix multiplication workloads of the Transformer models incorporate huge input matrices, which are not feasible to be processed all at once. 
To address this, block matrix multiplication is used by partitioning the matrices into smaller submatrices (tiles) \cite{golub2013matrix}. The multiplication of the entire matrices is reformulated as a series of smaller block-wise multiplications, followed by summations of intermediate results. 
This technique facilitates the mapping of large-scale matrix multiplications onto systolic arrays by decomposing the computation into independent blocks that align with the dimensions of the systolic array. Also, it enhances data locality and optimizes cache utilization by operating on sub-blocks.
Algorithm \ref{alg:matrix_multiplication} shows how block matrix multiplication is processed. It partitions matrices $A$ and $B$ into $T \times T$ tiles, enabling block-wise multiplication. The innermost loops perform matrix multiplication and accumulations over tile regions, while the tiles remain in memory for as long as possible.

\begin{algorithm}[t]
\caption{Block (Tiled) Matrix Multiplication}
\label{alg:matrix_multiplication}
\begin{algorithmic}[1]
\REQUIRE Matrices $A \in \mathbb{R}^{M \times K}$, $B \in \mathbb{R}^{K \times N}$, Tile size $T$
\ENSURE Output matrix $C \in \mathbb{R}^{M \times N}$
\FOR{$j = 0$ to $N-1$ step $T$}
    \FOR{$k = 0$ to $K-1$ step $T$}
        \FOR{$i = 0$ to $M-1$ step $T$}
            \FOR{$ii = i$ to $\min(i+T-1, M-1)$}
                \FOR{$jj = j$ to $\min(j+T-1, N-1)$}
                    \FOR{$kk = k$ to $\min(k+T-1, K-1)$}
                        \STATE $C[ii][jj] \mathrel{+}= A[ii][kk] \times B[kk][jj]$
                    \ENDFOR
                \ENDFOR
            \ENDFOR
        \ENDFOR
    \ENDFOR
\ENDFOR
\end{algorithmic}
\end{algorithm}

\section{Reconfigurable PE Architecture}

Multiplication decomposition is applied by spatially distributing the workload across multiple MAC units or by temporal scheduling.
Operands are divided into parallel subwords or multi-bit serial operands. 
For parallel subwords approach, the first operand remains at full-precision and is shared across all subword computations, while the second operand is partitioned into parallel subwords, achieving higher throughput for reduced-precision computations.
For multi-bit serial approach, both operands are partitioned into parallel subwords, and the full-precision result is spatially and temporally obtained, depending on width and number of instantiated MAC units.
To select the optimal reconfigurable PE architecture, multiple trade-offs are studied, including the number and width of instantiated MAC units, spatial-temporal mapping, and operands scheduling.

On the other hand, matrix multiplication stages in attention workloads of Transformer-based models are categorized into activation-to-weight and activation-to-activation workloads. 
Activation-to-weight workloads utilize 60\%–80\% of the total attention workload, such as Q, K, V, and attention output projections \cite{kim2023full}.   
Activation-to-activation workloads represent attention scores and attention output.
The majority of quantization frameworks reduce model weights to lower-precision and keep activations at full-precision to avoid performance degradation.
Collectively, the optimal reconfigurable PE design adapts to different precisions at run-time for quantized Transformer workloads, and maximizes throughput, memory efficiency, and energy gains.

\begin{figure}[b] 
  \centering
  \includegraphics[width=\linewidth, keepaspectratio]{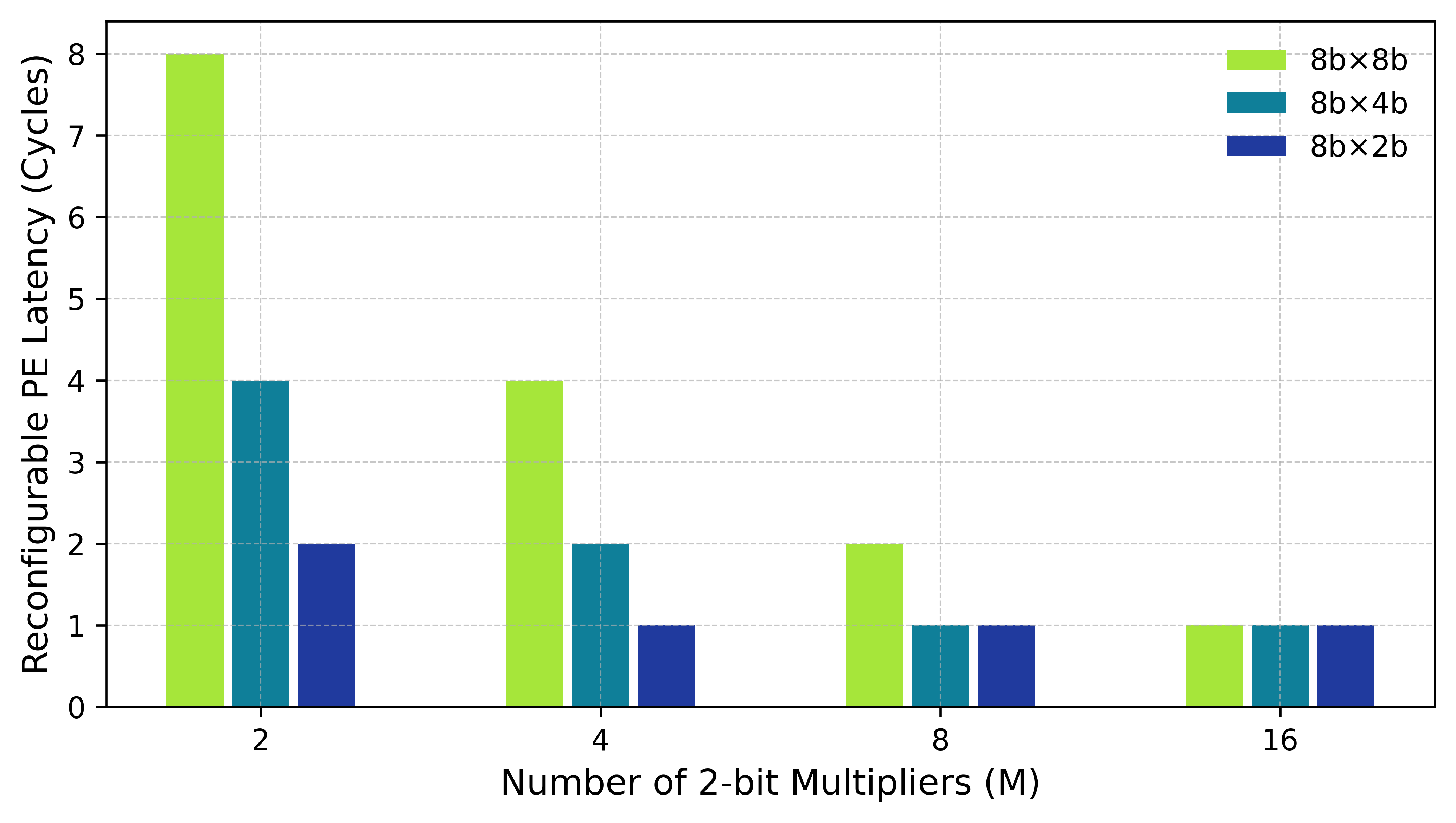} 
  \caption{Latency modeling results for the proposed reconfigurable PE across varying numbers of 2-bit multipliers ($M$ = 2, 4, 8, 16), where each bar represents one of the operand configurations: 8b×8b, 8b×4b, and 8b×2b.}
  \label{Fig.2}
\end{figure}

The proposed reconfigurable PE is similar to D\&C architecture with a set of 2-bit multipliers, accumulators, and registers.
Contrary to D\&C MAC units, the shifters and accumulators are shared across each PE column to save the area and power consumption associated with these units.
The latency analysis for the proposed reconfigurable PE, across varying numbers of 2-bit multipliers ($M$ = 2, 4, 8, 16), is shown in Fig. \ref{Fig.2} based on \eqref{eq:1}. 
The operand configurations of 8b×8b, 8b×4b, and 8b×2b are considered for the evaluation. 
It is shown that the latency decreases proportionally with increasing $M$, due to increased parallelism. 
The latency gap narrows as $M$ increases between operand configurations, reaching one cycle  at $M$=16. 
Particularly, for 8b×8b configuration, the latency is gradually reduced by increasing the number of 2-bit multipliers, stabilizing at one cycle with 16 multipliers. 
For 8b×4b operations, the PE latency stabilizes at one cycle with 8 multipliers. 
As a result, the PE doubles the throughput when instantiating 16 multipliers. 
For 8b×2b configuration, PE latency stabilizes at one cycle with 4 multipliers, quadrupling the throughput by instantiating 16 multipliers.
As a result, the optimal design for the proposed reconfigurable PE to instantiate 16 2-bit multipliers.

{\small
\begin{equation}
    \mathit{Latency_{PE}}  =
     \left\lceil \frac{1}{M} \times \left(
        \frac{OW_{1st} \times OW_{2nd}}{MW^2}
    \right) \right\rceil
\tag{1}\label{eq:1}
\end{equation}
}

\noindent where the notations:
\begin{itemize}
    \item $M$: Number of 2-bit multipliers
    \item $MW$: Operand width of each multiplier 
    \item $OW_{1st}, OW_{2nd}$: Operand bitwidths (multiple of 2-bit) \\
\end{itemize}

\begin{figure*}[t] 
  \centering
  \includegraphics[width=\textwidth, keepaspectratio]{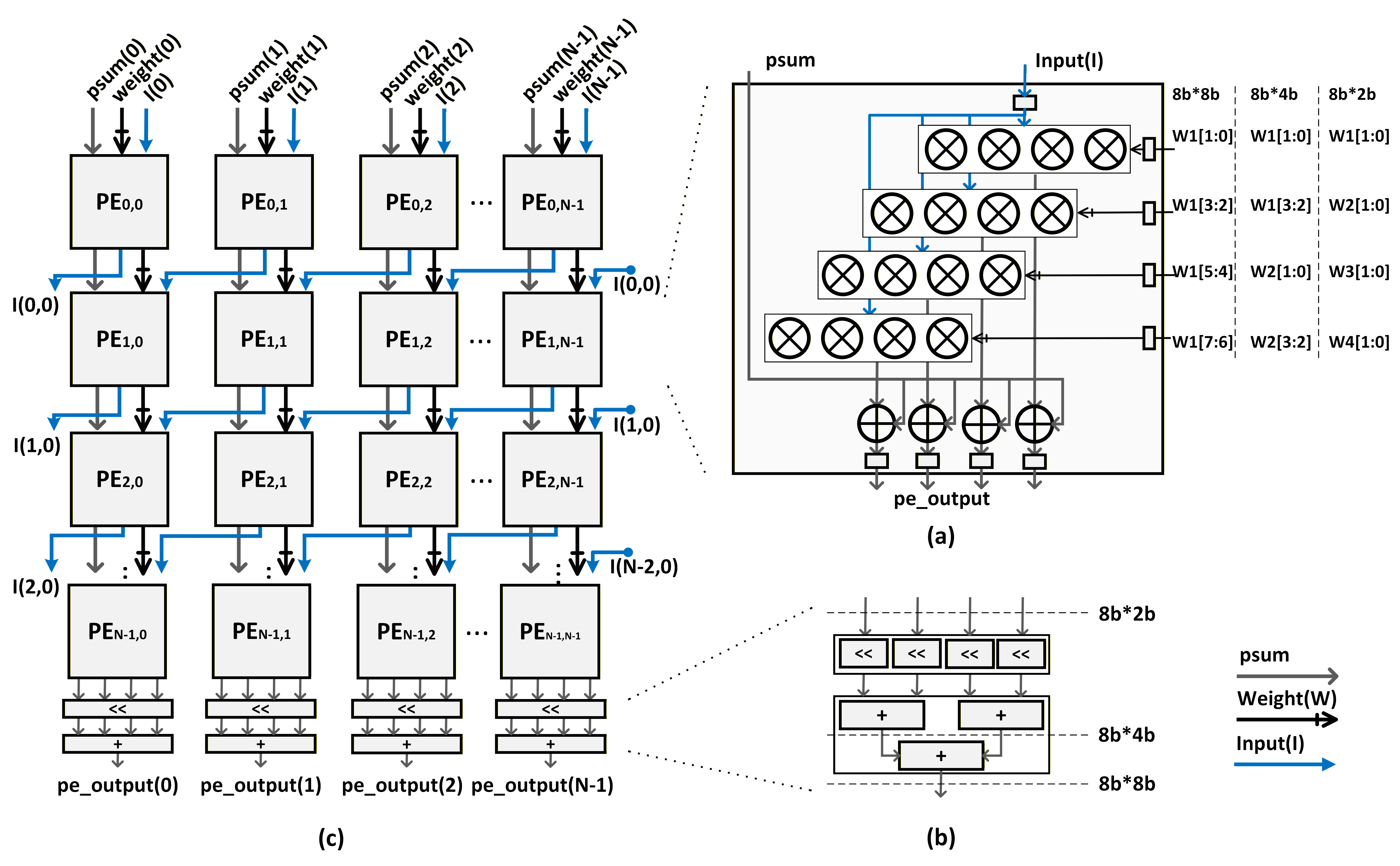} 
  \caption{
  (a) The proposed reconfigurable PE block diagram, consisting of 16 2-bit multipliers, arranged into four groups, four dedicated psums accumulators, along with enabled registers for weights, inputs and psums. (b) The reconfigurable unit of shared shifters and accumulators per each PE column adapting to different precisions (8b×8b, 8b×4b, and 8b×2b). (c) The proposed ADiP architecture comprising $N$×$N$ PEs, where inputs move diagonally across PE rows, transitioning from one row to the next, boundary PEs are diagonally connected, so that the registered inputs from the leftmost PE column feed into the inputs of the rightmost PE column in the subsequent row, weights are loaded vertically, and psums are accumulated vertically along the columns as well.}
  \label{Fig.3}
\end{figure*}

The proposed reconfigurable PE comprises 16 2-bit multipliers, group accumulators, and enabled registers for weights, inputs, and psums, as shown in Fig. \ref{Fig.3}(a). 
To obtain matrix multiplication, additional shifters and accumulators are required for each PE. However, these shifters and accumulators are shared across each PE column to reduce utilized area and energy consumption, as shown in Fig. \ref{Fig.3}(b).
The proposed reconfigurable PE adapts to different matrix multiplications with operand configurations of 8b×8b, 8b×4b, and 8b×2b.
For each matrix tile, the first operand (input activation) is common to all stationary subwords represented in 8-bit, 4-bit, or 2-bit.
The proposed reconfigurable PE balances between reconfigurability, area, and energy consumption overheads by selecting a few operating modes that cover the common precisions in quantized Transformer models. 
In addition, each PE incorporates four psums accumulators along with four fused and pipelined buses. 
The weight register stores a single weight value for single matrix multiplication represented in 8-bit, or 2 and 4 interleaved values encoded in 4-bit or 2-bit for multi-matrix multiplication, respectively. The input is registered and then shared with four groups of 2-bit multipliers.

\section{ADiP Architecture}

ADiP is an adaptive-precision systolic array architecture for matrix multiplication acceleration in quantized Transformer-based models. 
The proposed architecture incorporates $N$ × $N$ reconfigurable PEs along with shared shifters and accumulators across each PE column, as shown in Fig. \ref{Fig.3}(c). 
ADiP adapts to different precisions, and eliminates input and output synchronization FIFOs required by conventional WS architecture. 
The proposed architecture allows symmetric single-matrix multiplication or multi-matrix multiplication with asymmetric operands and shared input matrix.
In addition, ADiP architecture adapts to different precisions such as 8b×8b, 8b×4b, and 8b×2b, achieving higher throughput and memory efficiency by up to 4×. 
Moreover, ADiP improves data reuse and PE utilization by keeping the weight matrix stationary and sharing input activation matrix for asymmetric matrix multiplication.
Weights are loaded vertically, and partial summations (psums) are accumulated vertically along PE columns as well. 
Additionally, boundary PEs are diagonally connected, with the input registers of the leftmost PE column are connected to the inputs of the rightmost PE column in the next row. 
The activations move diagonally, passing from one PE row to the next.

As discussed earlier, the proposed reconfigurable PE consists of 16 2-bit multipliers, arranged into four groups with their internal accumulators. 
Based on the analytical modeling, the PE is configured to process 8b×8b multiplications in one cycle. 
Correspondingly, the throughput is doubled and quadrupled for 8b×4b and 8b×2b operations, respectively.
In addition, each PE instantiates four psum accumulators, and enabled registers for weights, input and psums. Psums are grouped and pipelined into four fused buses. 
In addition, the unit of shifters and accumulators is shared across each PE column instead of incorporating dedicated unit in each PE, as shown in Fig. \ref{Fig.3}(b).
This unit is reconfigured for different precision modes, such as 8b×8b, 8b×4b, and 8b×2b. 
The psum output of last PE row is passed to the shared shifters, and the shifters output is passed to shared accumulators. 
The output is selected directly from the last PE output, first stage, or second stage of accumulators, depending on the precision mode.

\begin{figure*}[t] 
  \centering
  \includegraphics[width=0.9\textwidth, keepaspectratio]{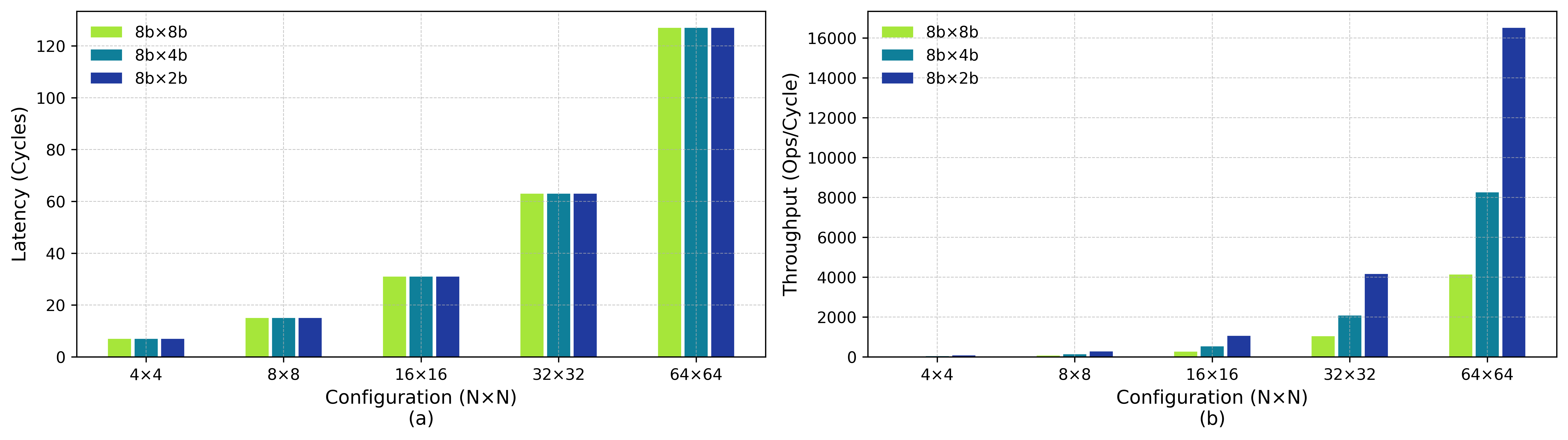} 
  \caption{Analytical modeling results for ADiP architecture in terms of ADiP latency and throughput with 16 multipliers per each reconfigurable PE ($M$ = 16) across different configurations for ADiP architecture.}
  \label{Fig.4}
\end{figure*}

\subsection{ADiP Analytical Modeling}

The analytical models for ADiP are developed to evaluate their latency and throughput compared to DiP models.
The latency model for ADiP considers the computation across all $N$ PE rows, number of MAC pipeline stages $S$, and external shift/add delays $E$, as shown in \eqref{eq:2}. 
The latency model is generic for any architecture configuration of \textit{N × N}, number of 2-bit multipliers $M$, operand widths, and pipeline delays. 
Moreover, the throughput for ADiP is calculated as the total number of operations (multiplications and additions) divided by ADiP latency, as derived by \eqref{eq:3}.
The analytical models provide a basis for the architectural trade-off under different architectural configurations.

\vspace{2mm}

$\mathit{Latency_{ADiP}}$ =
{\small
\begin{equation}
    N \times \left\lceil \frac{1}{M} \times \left(
        \frac{OW_{1st} \times OW_{2nd}}{MW^2}
    \right) \right\rceil + N + S + E - 2
\tag{2}\label{eq:2}
\end{equation}
}

$\mathit{Throughput_{ADiP}}$ =
{\small
\begin{equation}
    \frac{
        2 \times \left\lceil M \times \frac{(MW)^2}{OW_{1st} \times OW_{2nd}} \right\rceil \times N^3
    }{
        N \times \left\lceil \frac{1}{M} \times \frac{OW_{1st} \times OW_{2nd}}{MW^2} \right\rceil + N + S + E - 2
    }
\tag{3}\label{eq:3}
\end{equation}
}

\noindent where the notations:
\begin{itemize}
    \item $N$: Number of PEs per ADiP's row/column
    \item $S$: Number of MAC pipeline stages
    \item $E$: Number of external shift/add stages
\end{itemize}

Fig. \ref{Fig.4} shows a comparative analysis ADiP latency and throughput, across varying architecture configurations ($N$ = 4, 8, 16, 32, 64), based on analytical modeling derived by \eqref{eq:2} and \eqref{eq:3}. 
The evaluation considers operand configurations of 8b×8b, 8b×4b, and 8b×2b. 
ADiP latency results follow the same trend as same as reconfigurable PE latency, as shown in Fig. \ref{Fig.4}(a). 
The proposed ADiP offers throughput advantages, achieving up to 4× higher throughput, as shown in Fig. \ref{Fig.4}(b). 
By increasing the size of ADiP architecture, the throughput increases linearly, especially for quantized operations, such as 8b×4b and 8b×2b.

\begin{figure*}[b] 
  \centering
  \includegraphics[width=\textwidth, keepaspectratio]{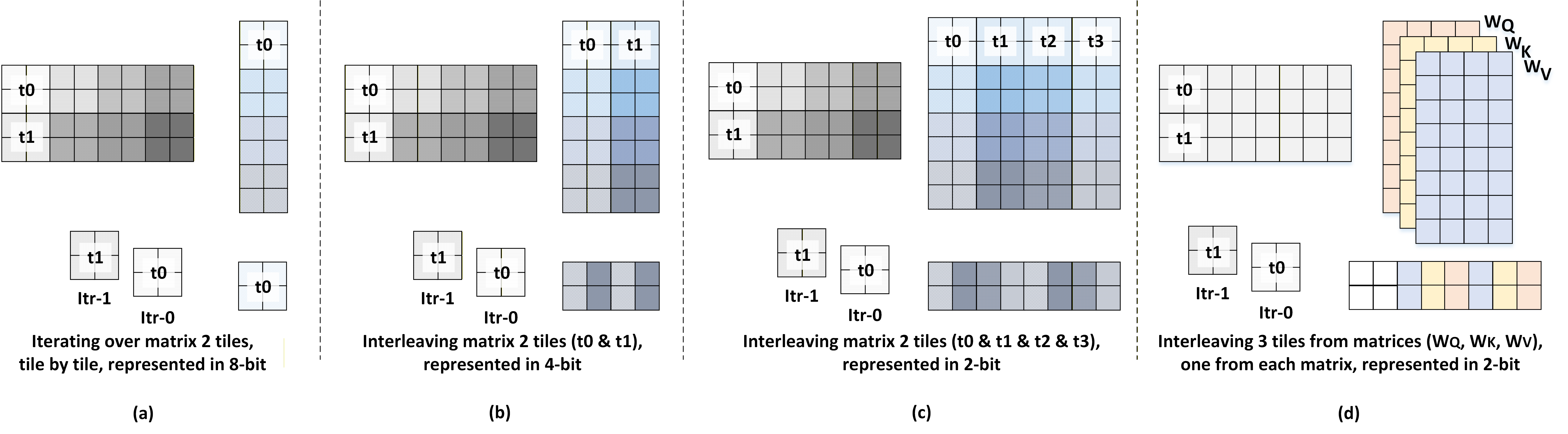} 
  \caption{An example of interleaving process in ADiP for different matrix multiplications: (a) 8b×8b multiplication using a single tile without interleaving. (b) 8b×4b multiplication with two interleaved weight tiles, where every two adjacent elements are converted to 4-bit precision. (c) 8b×2b multiplication with four interleaved weight tiles, where every four adjacent elements are converted to 2-bit. (d) Q, K, and V projections (8b×2b) in attention layers, where three weight tiles (one from each matrix) are interleaved, with every three adjacent elements converted to 2-bit.}
  \label{Fig.5}
\end{figure*}

\subsection{ADiP Dataflow}

\begin{figure}[t]
  \centering
  \includegraphics[width=\linewidth, keepaspectratio]{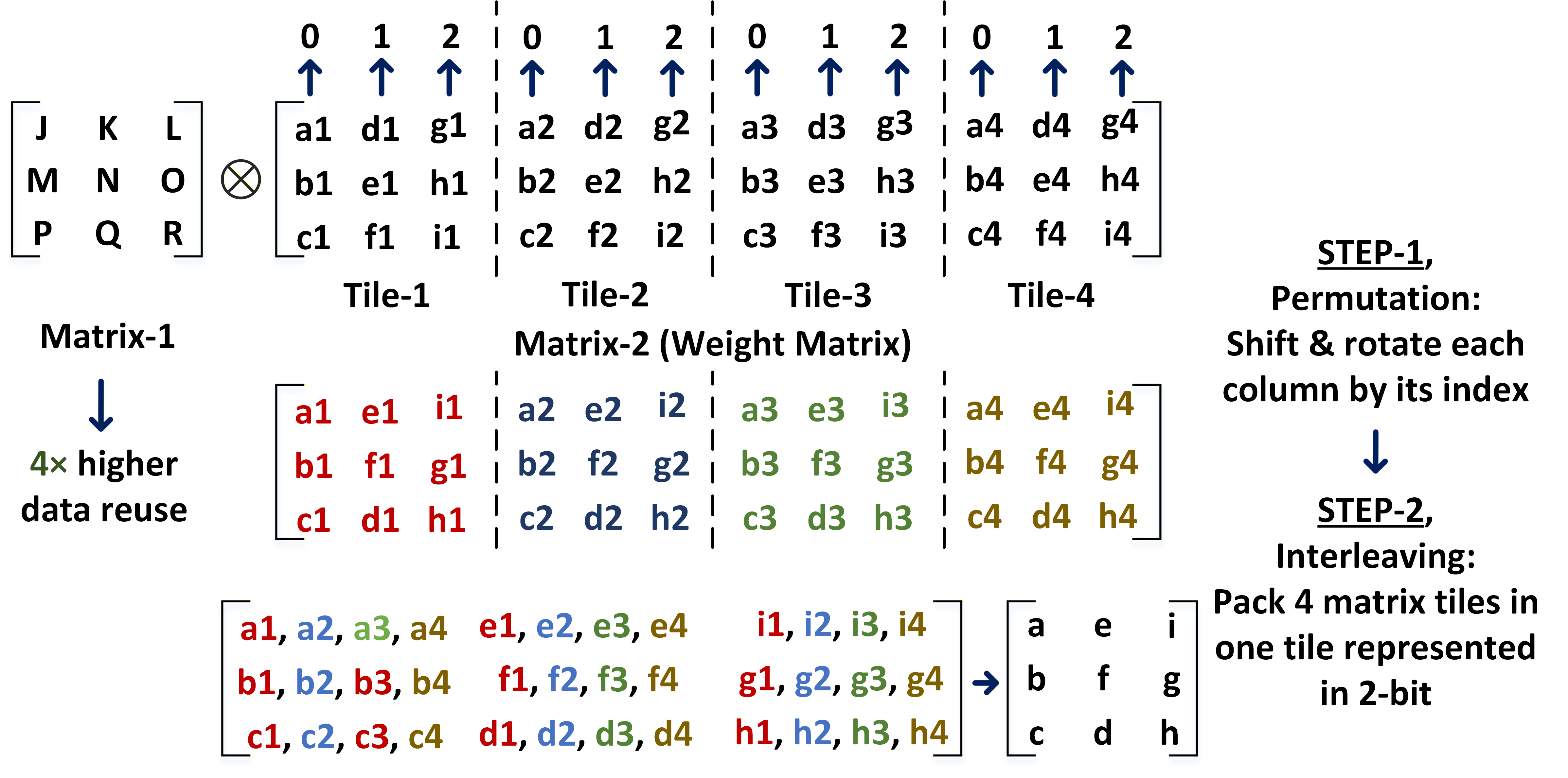} 
  \caption{An example of weights preparation for 8b×2b matrix multiplication on the ADiP architecture, involving permutation and interleaving four weight tiles, with every group of four adjacent elements converted to 2-bit and concatenated.}
  \label{Fig.6}
\end{figure}

The proposed ADiP architecture adopts a dataflow for mapping input matrices to the architecture.
This dataflow enables eliminating input and output synchronization FIFOs, and process multiple matrix multiplication workloads with the same input activation matrix.
The preprocessing involves two steps: weight permutation, and then interleaving.
The permutation step is similar to DiP dataflow, where each matrix tile is permutated by rotating each column upward by its column index \cite{abdelmaksoud2025dip}.
Then, the dataflow perform the interleaving step for weight matrices in case of 8b×4b and 8b×2b matrix multiplications. 
For 8b×8b operations, no interleaving is required as only single stationary matrix tile is involved, as shown in Fig. \ref{Fig.5}(a). 
For 8b×4b operations, two distinct weight tiles are interleaved, with every pair of adjacent elements converted to 4-bit and concatenated, as shown in Fig. \ref{Fig.5}(b). 
Similarly, for 8b×2b operations, four weight tiles are interleaved, with every group of four adjacent elements converted to 2-bit and concatenated, as shown in Fig. \ref{Fig.5}(c). 
Fig. \ref{Fig.5}(d) shows another 8b×2b matrix multiplication variation for multiplying input activation matrix by three distinct matrices of queries, keys, and values.
In this case, three weight tiles (one from each matrix) are interleaved, with every triplet of adjacent elements converted to 2-bit and concatenated.
This case is important when the core utilization is limited by the ratio between the head size and the ADiP core size, which may otherwise lead to under-utilization.
Fig. \ref{Fig.6} shows an example of weights preparation for 8b×2b matrix multiplication for the ADiP architecture. 
The preprocessing involves two steps, including permutation, then interleaving four weight tiles, with every group of four adjacent elements converted to 2-bit and concatenated. 
As a result, this enables ADiP architecture to process multi-matrix multiplication with asymmetric operands and a shared input matrix, achieving up to 4× higher data reuse.
The preprocessing is performed offline in case of activation-to-weight workloads, and at runtime in case of activation-to-activation workloads by efficiently re-scheduling memory access across multi-bank memories with almost zero overhead.

\section{Evaluation \& Results}

This section presents the hardware design space exploration for the proposed ADiP architecture, followed by the evaluation using attention workloads. Finally, ADiP is compared with the state-of-the-art accelerators.

\begin{table*}[t]
\caption{Area, Power Overhead, Total Overheads and Throughput Gain for ADiP architecture compared to DiP architecture}
\label{tab:overheads_throughput}
\centering
\resizebox{\linewidth}{!}{%
\begin{tabular}{ccccccc}
\hline
\textbf{Size} & \textbf{Area Overhead (×)} & \textbf{Power Consumption Overhead (×)} & \textbf{Total Overhead (×)} & \multicolumn{3}{c}{\textbf{Throughput Gain (×)}} \\ \cline{5-7} 
 &  &  &  & \textbf{8b×8b} & \textbf{8b×4b} & \textbf{8b×2b} \\ \hline
4×4   & 1.41 & 1.63 & 2.3 & 1 & 2 & 4 \\
8×8   & 1.34 & 1.59 & 2.13 & 1 & 2 & 4 \\
16×16 & 1.27 & 1.57 & 1.99 & 1 & 2 & 4 \\
32×32 & 1.29 & 1.63 & 2.1 & 1 & 2 & 4 \\
64×64 & 1.3 & 1.69 & 2.2 & 1 & 2 & 4 \\ \hline
\end{tabular}%
}
\end{table*}

\subsection{Hardware Design Space Exploration}

\begin{figure}[t]
  \centering
  \includegraphics[width=\linewidth, keepaspectratio]{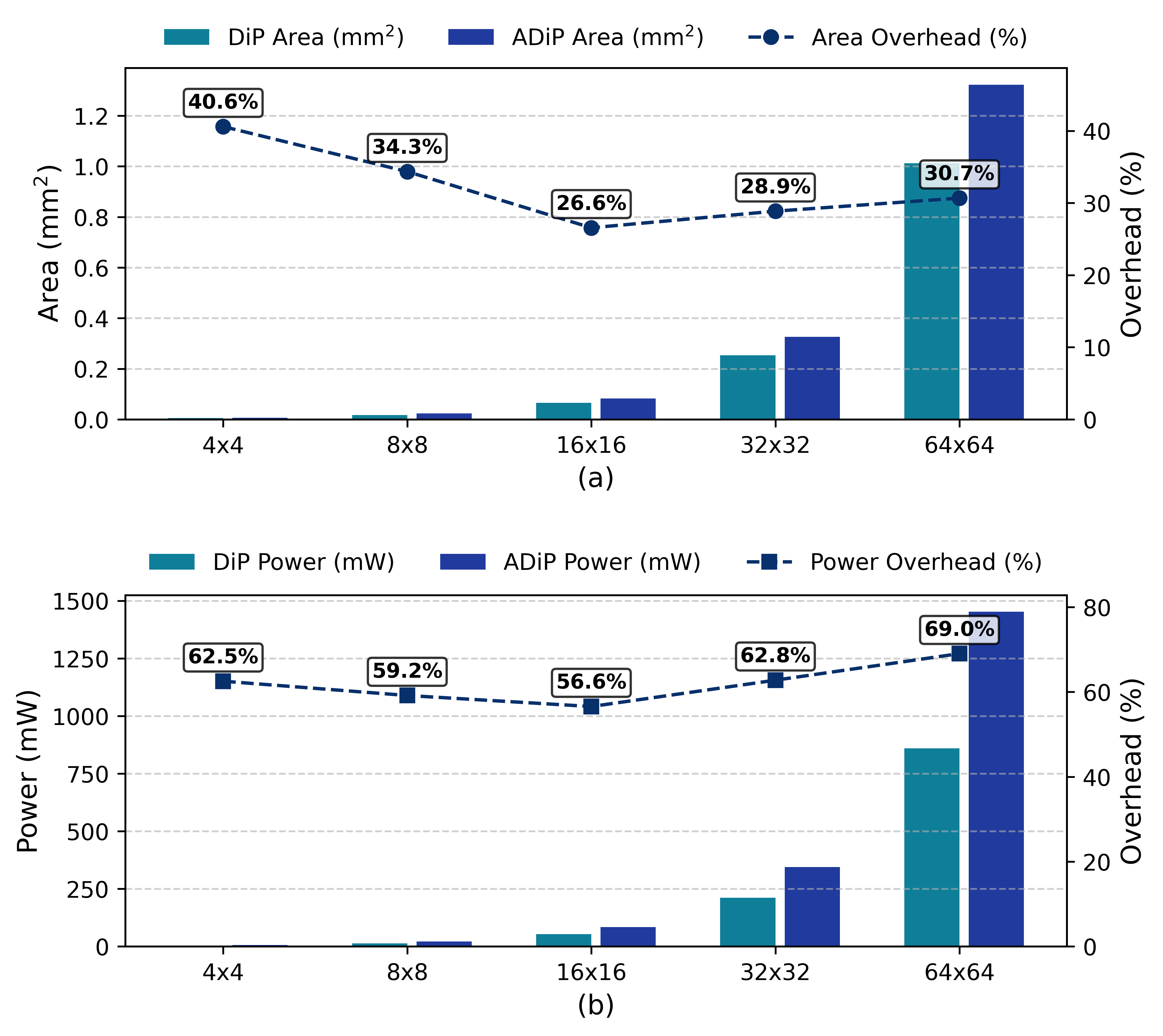} 
  \caption{Area and power consumption comparison of DiP and ADiP architectures across array sizes from 4×4 to 64×64. (a) Area breakdown for both architectures. (b) Power consumption breakdown for both architectures.}
  \label{Fig.7}
\end{figure}

Hardware design space exploration is developed for DiP and ADiP architectures at different sizes. Both architectures are scaled from 4×4 to 64×64 with variants: 4×4, 8×8, 16×16, 32×32, and 64×64. A parameterized HDL design using Verilog is developed. All designs are implemented from synthesis to GDSII using Cadence Genus and Innovus tools with commercial 22nm technology with operating voltage 0.8 V at a frequency of 1 GHz.
Table \ref{tab:overheads_throughput} demonstrates that the throughput gains scale linearly for the proposed ADiP architecture compared to DiP architecture, achieving 2×, and 4× for 8b×4b, and 8b×2b configurations, respectively. The gains achieved are obtained by involving reconfigurable PEs in the proposed ADiP architecture compared to conventional PEs with INT8 MAC units in the DiP architecture.
The throughput gain comes with area and power consumption overheads for ADiP architecture compared to DiP architecture, as presented in Table \ref{tab:overheads_throughput} and Fig. \ref{Fig.7}. 
The total overhead is between 1.99× and 2.3×. 
In particular, as reported by Cadence Innovus, the total overhead is reduced from 2.3× at 4×4 to 1.99× at 16×16, and slightly increases to 2.1× and 2.2× at 32×32 and 64×64 configurations, respectively. As shown in Fig. \ref{Fig.7}(a), area overhead decreases from 40.6\% at 4×4 to 26.6\% at 16×16, and further stabilizes at 28.9\% and 30.7\% for 32×32 and 64×64 configurations, respectively, due to shared accumulators that improve resource utilization at larger scales. In addition, power consumption overhead in Fig. \ref{Fig.7}(b) drops from 62.5\% at 4×4 to 56.6\% at 16×16, and further stabilizes at 62.8\% and 69\% for 32×32 and 64×64 configurations, respectively.

\begin{figure}[t]
  \centering
  \includegraphics[width=\linewidth, keepaspectratio]{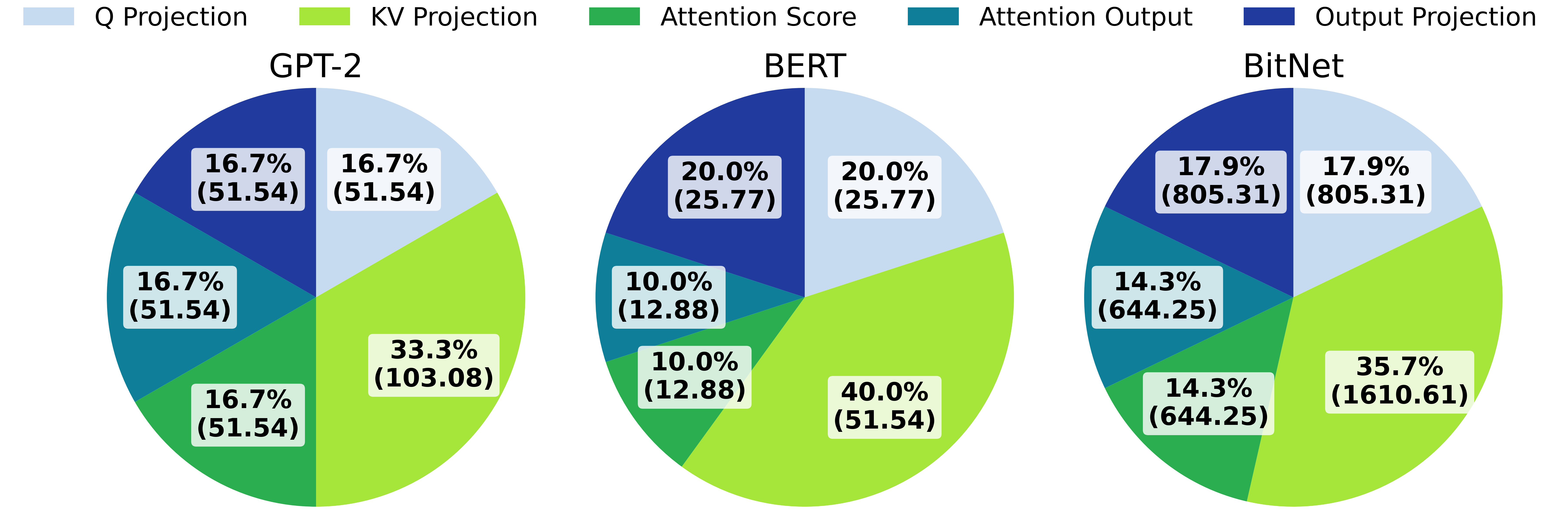} 
  \caption{Attention workloads breakdown for the evaluated Transformer-based models, including GPT-2 medium, BERT large, and BitNet-1.58B.}
  \label{Fig.8}
\end{figure}

\begin{figure*}[b]
  \centering
  \includegraphics[width=\textwidth, keepaspectratio]{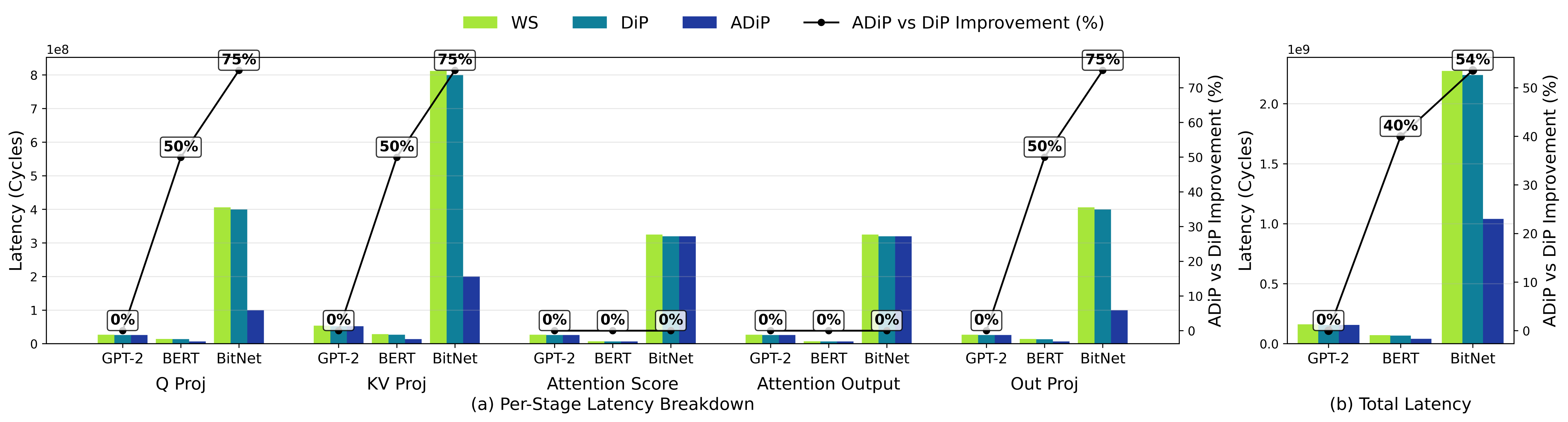} 
  \caption{Latency comparison between WS, DiP, and ADiP architectures using attention workloads from Transformer-based models: GPT-2 medium, quantized BERT large (4-bit), and BitNet-1.58B. (a) Per-stage latency analysis. (b) Total latency analysis with improvement annotations comparing ADiP to DiP for per-stage and total latency comparisons.}
  \label{Fig.9}
\end{figure*}

\begin{figure*}[b]
  \centering
  \includegraphics[width=\textwidth, keepaspectratio]{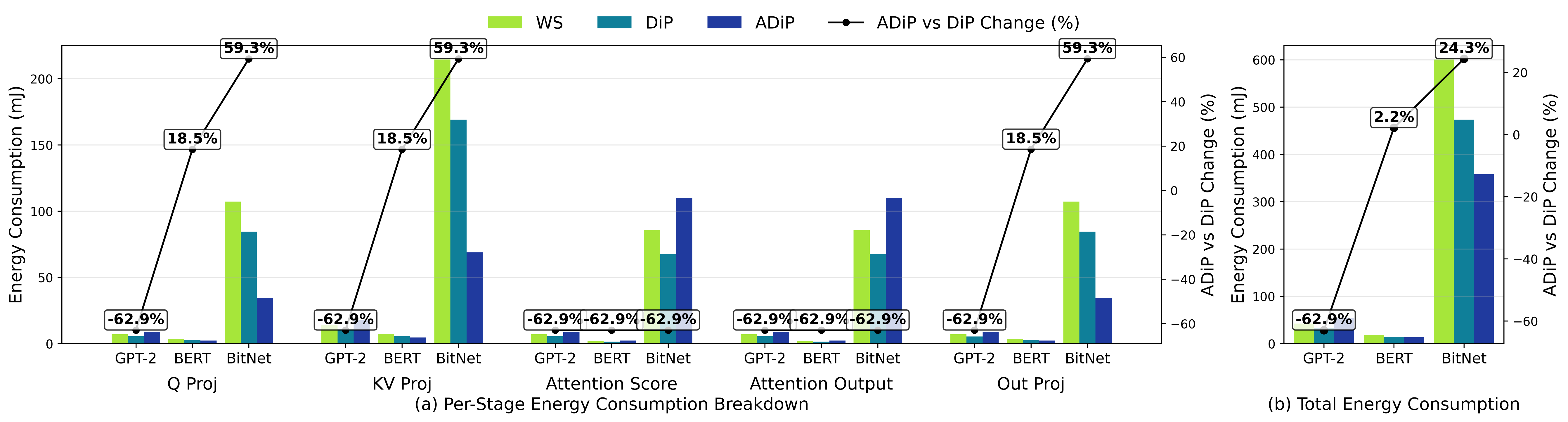} 
  \caption{Energy consumption between WS, DiP, and ADiP architectures using attention workloads from Transformer-based models: GPT-2 medium, quantized BERT large (4-bit), and BitNet-1.58B. (a) Per-stage energy consumption analysis. (b) Total energy consumption analysis with annotations for energy change comparing ADiP to DiP, where negative values indicate energy overhead.}
  \label{Fig.10}
\end{figure*}

\subsection{Evaluation Using Attention Workloads}

ADiP is evaluated using matrix multiplication workloads derived from the attention layers of three models: GPT-2 medium \cite{radford2019language}, BERT large \cite{devlin2019bert}, and BitNet-1.58B \cite{diao2023bitnet, wang20241}.
GPT-2 medium and BERT large both feature 24 layers, a hidden size of 1024, and 16 attention heads with a head dimension of 64. GPT-2 medium is a decoder-only model with a maximum sequence length of 1024 and operates at 8-bit precision, while BERT large is an encoder-only model supporting sequences up to 512 tokens and operates at 4-bit precision. 
The attention (MHA) workload size for the GPT-2 medium and BERT large models is nearly 309.24 GOPS, and 128.85 GOPS, respectively. In contrast, BitNet-1.58B is a significantly larger decoder model with 30 layers, a hidden size of 2560, 20 attention heads (each with dimension 128), and a maximum sequence length of 2048. It employs 2-bit weights, resulting in a compact architecture with a model size of approximately 2.5 billion parameters and attention workload size of nearly 4.51 TOPS. 
The per-stage breakdown is presented in Fig. \ref{Fig.8}, showing that the projection (activation-to-weight) workloads utilize from 60\% to 80\% from the total workload.
A cycle-accurate simulator is developed to evaluate the latency, energy consumption, and memory access for WS, DiP, and ADiP architectures.
The simulator employs analytical models for WS and DiP architectures, derived from the DiP work \cite{abdelmaksoud2025dip}.
In addition, the matrix multiplication workloads for each model are evaluated independently on the target architectures.

Fig. \ref{Fig.9} presents a comparative analysis of latency and energy consumption between WS, DiP architectures, and the proposed ADiP architecture across different attention stages (Q, K, V, and output projections) and activation-to-activation workloads (attention scores and attention output) from the evaluated models. 
Activation-to-activation workloads are performed in 8b×8b, while projection workloads are performed in 8b×8b, 8b×4b, and 8b×2b for GPT-2 medium, BERT large, and BitNet-1.58B, respectively. The evaluation is studied for WS, DiP and ADiP with size 32×32 to be fully-utilized during the processing of the evaluated attention workloads.
The per-stage breakdown reinforces the workload distribution of the attention workloads for each model presented in Fig. \ref{Fig.8}.
In addition, the annotations represent the improvement percentages, which reveals the improvement for each workload.
The obtained results reveal that ADiP achieves substantial latency improvements exclusively in the projection stages, showing 50\% and 75\% improvements, with total latency improvements of 40\% and 53.6\% for BERT large and BitNet-1.58B models compared to DiP, respectively. For GPT-2 medium with 8-bit precision, it incurs no latency overhead.
This selective improvement occurs because projection stages involve activation-to-weight multiplications that can effectively exploit ADiP's adaptive-precision capabilities through enhanced compute density mapping of low bitwidth operations. 
In contrast, activation-to-activation operations exhibit no latency benefits, as these stages involve dynamic data dependencies that cannot leverage quantization opportunities.

In the analysis of energy consumption shown in Fig. \ref{Fig.10}, ADiP exhibits differentiated behavior across projection and activation-to-activation workloads. For projection stages, including Q, K, V, and output projections, ADiP achieves energy efficiency benefits, particularly in quantized models such as BERT (4-bit) and BitNet-1.58B (2-bit). 
The reduced bitwidth operations in these models enable ADiP to execute more computations per cycle, leading to lower energy consumption. On the other hand, activation-to-activation workloads demonstrate an energy overhead for ADiP across all models. This overhead arises from the limited opportunity to exploit higher computational density in activation-to-activation stages. Notably, while the energy overhead for activation-to-activation stages is marginal, the projection stages compensate for this, resulting in a net positive energy savings for quantized workloads.
Figure \ref{Fig.10}(b) shows a total energy improvement of 2.3\% BERT large (4-bit) and 24.4\% BitNet-1.58B (2-bit) compared to DiP, while the GPT-2 medium (8-bit) model shows an energy overhead of 62.8\%. 
These results show that ADiP's adaptive-precision capabilities are most advantageous for highly quantized attention workloads, where the computational acceleration outweighs the increased energy consumption overhead, making it particularly suitable for energy-efficient inference of ultra-low precision language models, such as BitNet model.

\begin{figure*}[t]
  \centering
  \includegraphics[width=\textwidth, keepaspectratio]{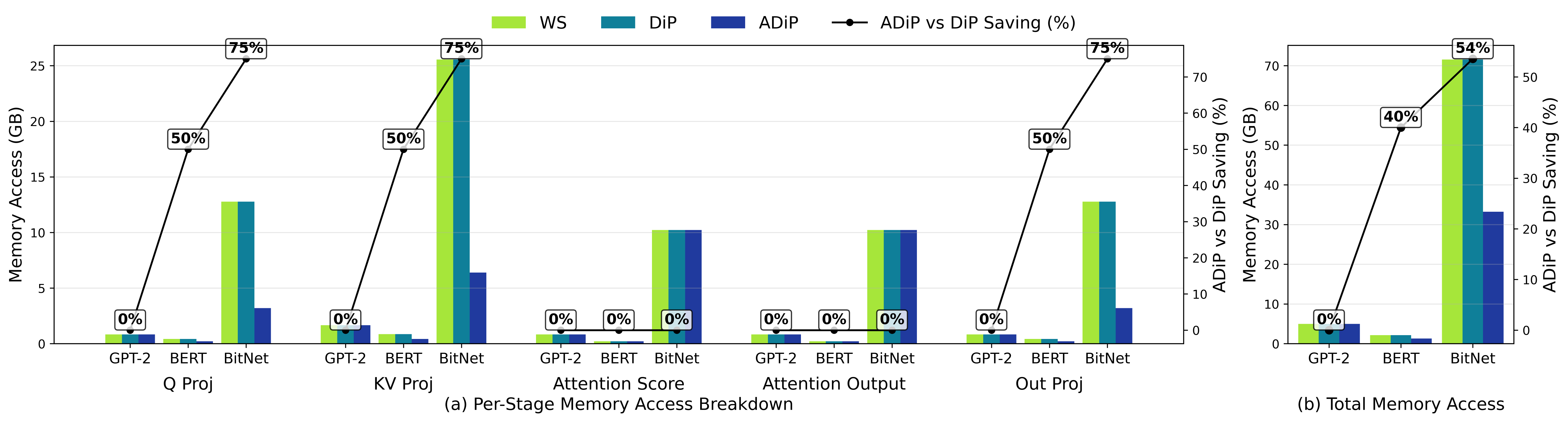} 
  \caption{Comparison of memory access between WS, DiP, and ADiP architectures using attention workloads from Transformer-based models: GPT-2 medium, quantized BERT large (4-bit), and BitNet-1.58B (2-bit). (a) Per-stage memory access analysis. (b) Total memory access analysis with improvement annotations comparing ADiP to DiP for per-stage and total memory access comparisons.}
  \label{Fig.11}
\end{figure*}

Finally, the results presented in Fig. \ref{Fig.11} illustrate the comparative analysis of memory accesses between WS, DiP, and ADiP architectures across GPT-2 medium (8-bit), BERT large (4-bit), and BitNet-1.58B (2-bit) workloads. The graph shows per-stage breakdown and the total memory access (in GB) for the evaluated workloads, while the annotations represent the relative memory access savings (\%) achieved by ADiP compared to DiP. 
ADiP achieves up to 4× higher memory efficiency, as one architectural benefits, for the quantized projection workloads by reducing the number of tile accesses of the first input matrix (represented in 8-bit), and adjusts the bit-width of memory accesses in the second input matrix tiles.  
For GPT-2 medium with 8-bit weights, the total memory accesses are the same. However, ADiP achieves a reduction of approximately 40.0\% for BERT large with 4-bit weights. For the large-scale BitNet-1.58B model with 2-bit weights, the savings further improve to 53.6\%, indicating that ADiP significantly reduces the memory accesses for highly quantized attention workloads. 

While DiP architecture outperforms conventional WS architecture in power consumption by up to 1.25×, area by up to 1.09×, and energy efficiency per area by up to 2.02×.
In addition, ADiP architecture outperforms DiP architecture in throughput and memory efficiency for quantized workloads.
These results confirm that ADiP achieves higher throughput and memory efficiency with minimal area and power consumption overheads at high compute densities compared to state-of-the-art architectures, making it well-suited for compute-intensive and memory-efficient applications.

\subsection{Comparison With State-Of-The-Art Accelerators}

Table \ref{table:literature_comparison} compares ADiP with DiP \cite{abdelmaksoud2025dip}, Google TPU V4i \cite{jouppi2021tpuv4i}, DTQAtten \cite{yang2022dtqatten}, DTATrans \cite{yang2023dtatrans}, and BitSystolic \cite{yangbitsystolic} highlighting ADiP's performance and area efficiency for matrix multiplication acceleration. 
Since each accelerator is implemented in different technology, the performance metrics are normalized to 22nm using DeepScaleTool \cite{sarangi2021deepscaletool, deepscaletool2025}. Area and energy efficiency results are presented before and after scaling to 22nm. 
ADiP architecture features 4,096 reconfigurable PEs in a 64×64 configuration, operating at 1 GHz using commercial 22nm technology. At a size of 64×64, ADiP offers peak throughput of 8.192 TOPS, 16.384 TOPS, and 32.768 TOPS for 8bx8b, 8bx4b, and 8bx2b operations, respectively. Furthermore, ADiP delivers energy efficiency with 5.64 TOPS/W, 11.28 TOPS/W, and 22.57 TOPS/W for 8bx8b, 8bx4b, and 8bx2b operations, respectively. In addition, it delivers an area efficiency (computational density) with 6.2 TOPS/mm², 12.41 TOPS/mm², and 24.82 TOPS/mm². 
The throughput and energy efficiency of DTQAtten and DTATrans are reported at 4b×4b precision, which is computationally equivalent to 8b×2b in terms of bit-serial operations, and thus no scaling adjustment is required for comparison.
However, BitSystolic reports peak throughput and energy efficiency metrics at 2b×2b operand precision. Since 8b×2b operations require a 4× increase in bit-serial compute cycles, both peak throughput and energy efficiency degrade by a corresponding factor of 4×.
These results demonstrate ADiP's capability to provide higher computational density for lower precision while minimizing energy consumption. Such metrics make ADiP particularly well-suited for highly-quantized Transformer-based applications.

\begin{table*}[t]
\centering
\caption{Comparison with State-of-The-Art Accelerators}
\label{table:literature_comparison}

\begin{tabular}{
>{\raggedright\arraybackslash}p{3.9cm}
>{\centering\arraybackslash}p{1.55cm}
>{\centering\arraybackslash}p{1.55cm}
>{\centering\arraybackslash}p{1.55cm}
>{\centering\arraybackslash}p{1.55cm}
>{\centering\arraybackslash}p{1.55cm}
>{\centering\arraybackslash}p{1.55cm}}

\toprule
& \textbf{\ ADiP \newline(this work)} & \textbf{\ \ DiP \newline \cite{abdelmaksoud2025dip}} & \textbf{Google TPU V4i\cite{jouppi2021tpuv4i}} & \textbf{\ \ BitSystolic \newline\cite{yangbitsystolic}} & \textbf{\ DTQAtten \newline\cite{yang2022dtqatten}} & \textbf{DTATrans \cite{yang2023dtatrans}}  \\

\midrule
\textbf{Architecture} &
64$\times$64 PEs &
64$\times$64 PEs &
4$\times$128$\times$128 PEs &
16$\times$16 PEs &
VSSA Modules &
VSSA Modules \\

\textbf{} &
Post-PnR &
Post-PnR &
Post-Silicon &
Post-Silicon &
Post-Syn &
Post-Syn \\

\textbf{Frequency (GHz)} &
1 &
1 &
1.05 &
1.5 &
1 &
1 \\

\textbf{Precision\textsuperscript{(1)}} &
A:8, W:2,4,8 &
A/W:8 &
A/W:8 &
A/W:2,4,8 &
A/W:4,8 &
A/W:4,8 \\

\textbf{Technology} &
22nm &
22nm &
7nm &
65nm &
40nm &
40nm \\

\textbf{Power (W)} &
1.452 &
0.858 &
175 &
0.0178 &
0.734 &
0.803 \\

\textbf{Area (mm\textsuperscript{2})} &
1.32 &
1 &
400 &
4 &
1.41 &
1.49 \\

\textbf{P. Throughput (TOPS)} &
32.768 \newline @ 8b×2b &
8.192 \newline @  8b×8b &
138 \newline @ 8b×8b &
0.403 \newline @ 2b×2b &
0.953 \newline @ 4b×4b &
1.304 \newline @ 4b×4b \\

\textbf{Area Efficiency (TOPS/mm²)} &
24.824 &
8.192 &
0.345 &
0.1 &
0.676 &
0.979 \\

\textbf{Energy Efficiency (TOPS/W)} &
22.567 &
9.548 &
0.786 &
26.7  &
1.298 &
1.623 \\

\textbf{Area Efficiency (TOPS/mm\textsuperscript{2})\textsuperscript{(2)}} &
24.824 &
8.192 &
0.017 &
0.935 &
2.302 &
2.984 \\

\textbf{Energy Efficiency (TOPS/W)\textsuperscript{(2)}} &
22.567 &
9.548 &
0.345 &
47.412 &
1.973 &
2.470 \\

\bottomrule
\end{tabular}

\begin{flushleft}
\footnotesize
\textsuperscript{(1)}Activation and weight precisions are denoted by A and W, respectively.

\textsuperscript{(2)}Power and area are normalized to 22nm using DeepScaleTool \cite{sarangi2021deepscaletool, deepscaletool2025}.
\end{flushleft}

\end{table*}

\vspace{-5mm}

\section{Conclusion}

This paper presents ADiP, an adaptive-precision systolic array architecture designed to efficiently accelerate matrix multiplication in quantized Transformers and LLMs. ADiP consists of $N$×$N$ reconfigurable PEs, along with shared shifters and accumulators. By supporting multiple precisions, ADiP significantly increases computational density and data reuse by up to 4×. Analytical models are developed to characterize latency and throughput across various configurations. In addition, a comprehensive hardware design space exploration is implemented in a commercial 22nm technology, showing up to 4× higher throughput with 2× area and power consumption overheads.
Furthermore, ADiP is evaluated on different attention workloads from GPT-2 medium, BERT large, and BitNet-1.58B models, delivering latency improvement of up to 53.6\%, and energy improvement of up to 24.4\% for BitNet-1.58B attention workloads. 
Additionally, ADiP achieves total memory access savings of 40\% and 53.6\% for the attention workloads of BERT large (4-bit) and BitNet-1.58B (2-bit), respectively.
At a configuration of 64×64 with 4,096 reconfigurable PEs, ADiP achieves a peak throughput of 8.192 TOPS, 16.384 TOPS, and 32.768 TOPS for 8b×8b, 8b×4b, and 8b×2b operations, respectively, making it suitable for quantized Transformer workloads.

\section*{Acknowledgment}
The authors would like to thank Dr Shady Agwa for his early recommendations.

\bibliographystyle{IEEEtran}
\bibliography{references}

\vspace{-40pt}
\begin{IEEEbiography}
[{\includegraphics[width=1in,height=1.2in,clip,keepaspectratio]{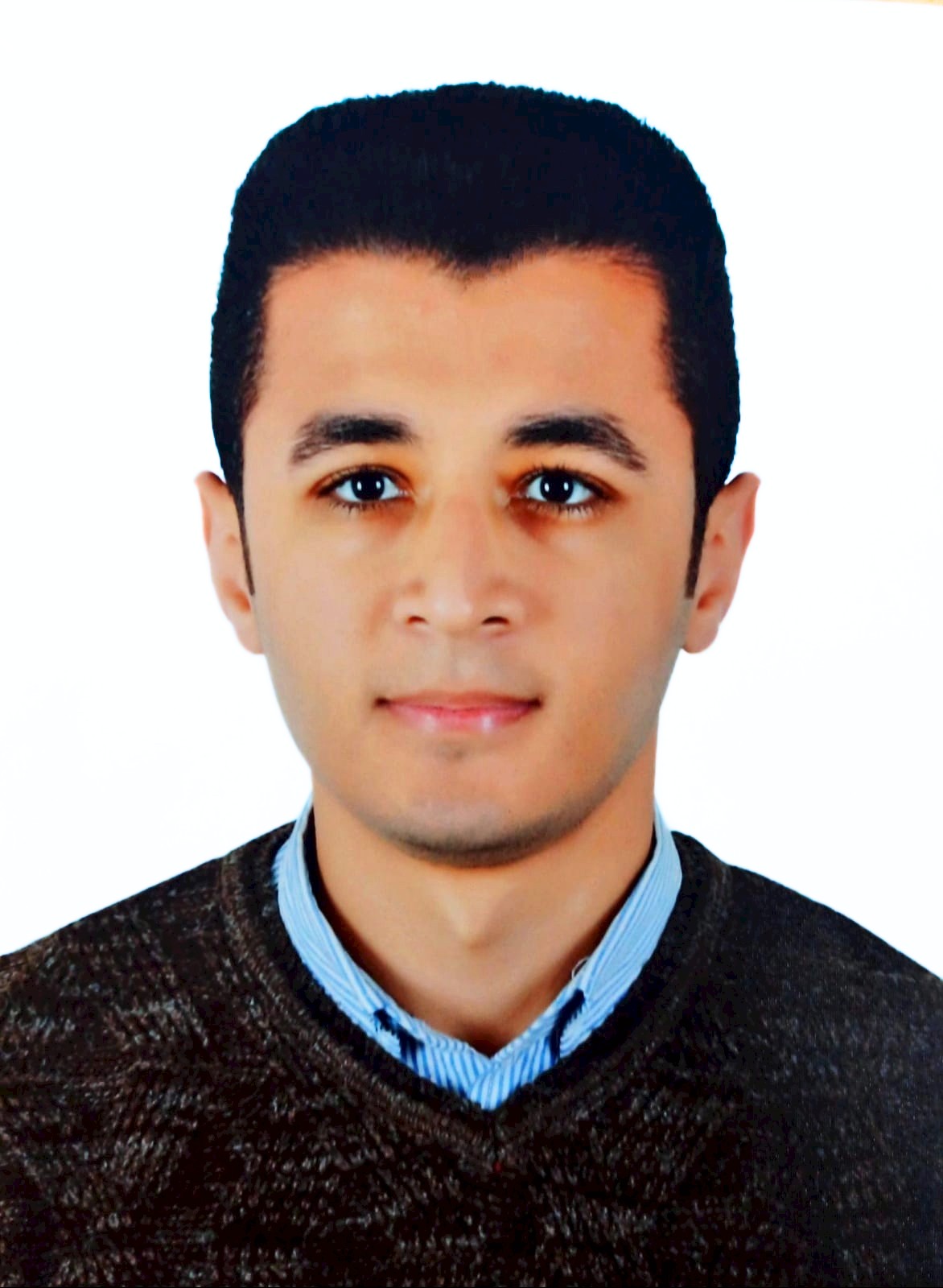}}]{Ahmed J. Abdelmaksoud}
is currently pursuing his PhD with the Centre for Electronics Frontiers (CEF) at the University of Edinburgh, UK. He received his BSc and MSc in Electronics Engineering from Cairo University, Egypt in 2018 and 2022, receptively. Since 2018, he has been actively involved in Digital ASIC design projects across both research and industry. His professional experience includes working as a Research Associate at the System-on-Chip Center, Khalifa University, UAE; an ASIC Physical Design Engineer at Si-Vision, Egypt; and a Research Assistant at the Opto-Nano Electronics Lab, Egypt. In addition, his current research interests primarily focus on developing spatial and specialized architectures for efficient AI hardware acceleration.
\end{IEEEbiography}
\vspace{-40pt}
\begin{IEEEbiography}
[{\includegraphics[width=1in,height=1.2in,clip,keepaspectratio]{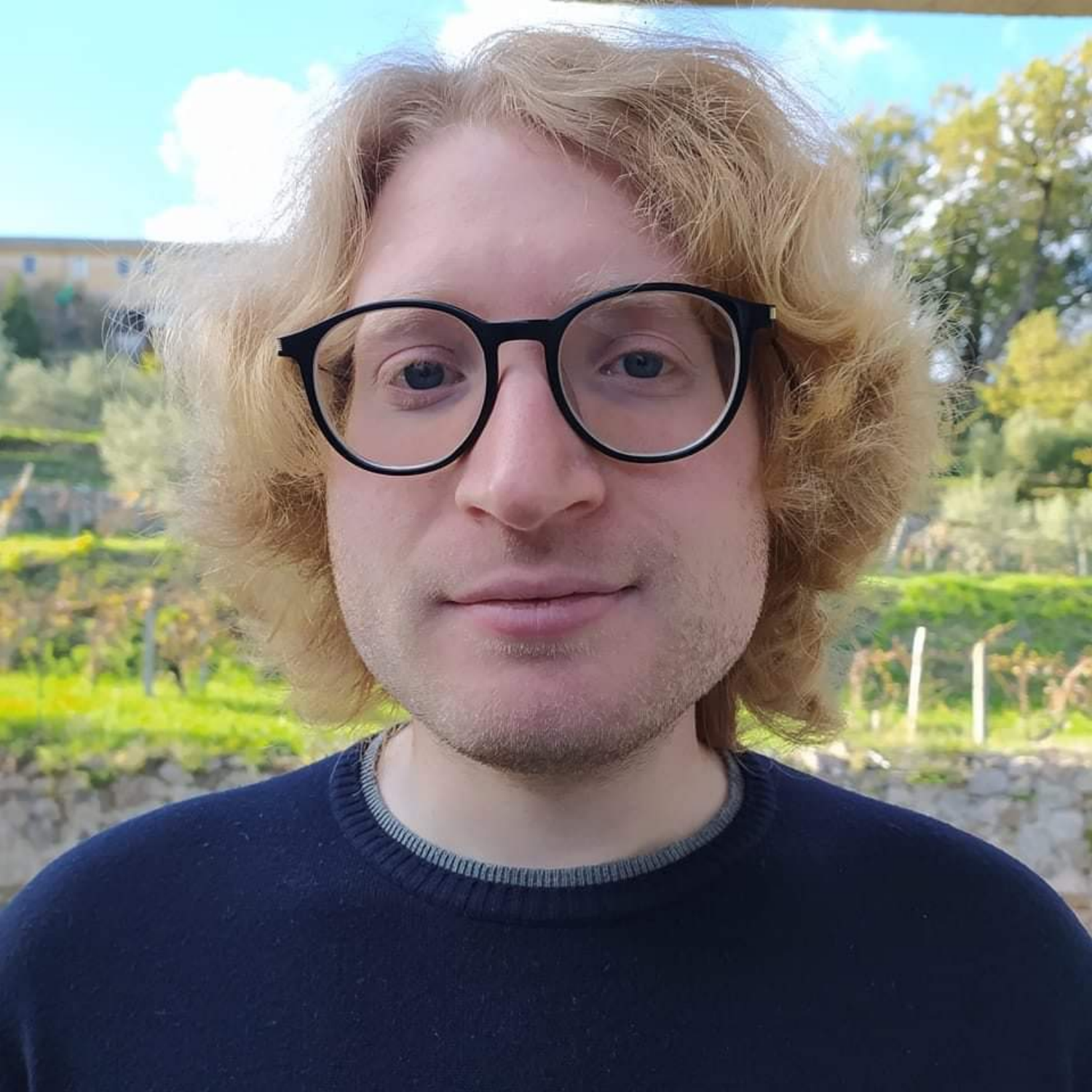}}]{Cristian Sestito}
(Member, IEEE) is a Research Fellow at the Centre for Electronics Frontiers CEF, The University of Edinburgh (UK). He received his BSc and MSc degree from University of Calabria (Italy), both in Electronic Engineering. He got his PhD in Information and Communication Technologies from the same university in 2023, focusing on Convolutional Neural Networks and their implementation on Field Programmable Gate Arrays (FPGA). In 2021/2022, Cristian was a Visiting Scholar at Heriot-Watt University, Edinburgh, working on neural networks’ compression. His research interests include digital design, embedded system design for AI on FPGA-based systems-on-chip, software simulators for neuromorphic AI, AI for circuits and systems design automation.
\end{IEEEbiography}
\vspace{-40pt}
\begin{IEEEbiography}
[{\includegraphics[width=1in,height=1.2in,clip,keepaspectratio]{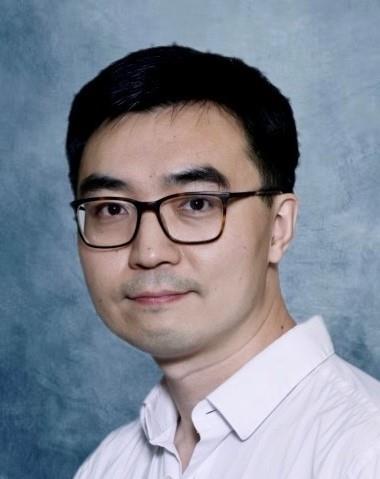}}]{Shiwei Wang} 
(Senior Member, IEEE) received the B.Eng. degree (Hons.) in electronic engineering from Zhejiang University, China, in 2010, and the Ph.D. degree in microelectronics from The University of Edinburgh, U.K., in 2014. He was a Research Assistant Professor at Shenzhen Institute of Advanced Technology, Chinese Academy of Sciences, China, from 2014 to 2015, a Senior Researcher at imec, Belgium, from 2015 to 2020, and an Associate Professor at the Department of Electronics and Computer Science, University of Southampton, U.K., from 2020 to 2022. In 2022, he joined the School of Engineering, The University of Edinburgh, where he is currently a Reader and a Bayes Innovation Fellow. His research interests include analog and mixed-signal integrated circuits for emerging application including implantable/wearable electronics, brain–machine interface, and sensor instrumentation.
\end{IEEEbiography}
\vspace{-40pt}
\begin{IEEEbiography}
[{\includegraphics[width=1in,height=1.2in,clip,keepaspectratio]{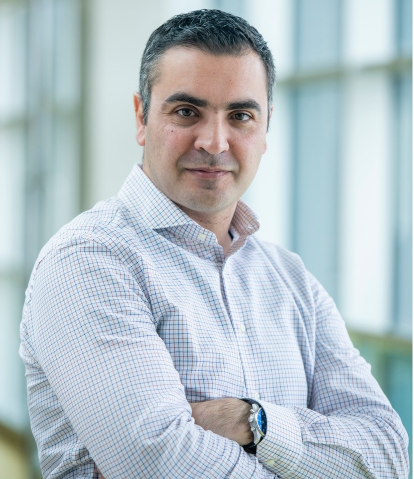}}]{Themis Prodromakis}
(Senior Member, IEEE) received the bachelor’s degree in electrical and electronic engineering from the University of Lincoln, U.K., the M.Sc. degree in microelectronics and telecommunications from the University of Liverpool, U.K., and the Ph.D. degree in electrical and electronic engineering from Imperial College London, U.K. He then held a Corrigan Fellowship in nanoscale technology and science with the Centre for Bio-Inspired Technology, Imperial College London, and a Lindemann Trust Visiting Fellowship with the Department of Electrical Engineering and Computer Sciences, University of California at Berkeley, USA. He was a Professor of nanotechnology at the University of Southampton, U.K. He holds the Regius Chair of Engineering at the University of Edinburgh and is Director of the Centre for Electronics Frontiers. He is currently a Royal Academy of Engineering Chair in emerging technologies and a Royal Society Industry Fellowship. His background is in electron devices and nanofabrication techniques. His current research interests include memristive technologies for advanced computing architectures and biomedical applications. He is a fellow of the Royal Society of Chemistry, the British Computer Society, the IET, and the Institute of Physics.
\end{IEEEbiography}

\end{document}